\documentclass[prd,showpacs,twocolumn]{revtex4}
\usepackage{amsmath}
\usepackage{amssymb}
\usepackage{epsfig}
\usepackage{graphicx}
\usepackage{array}
\usepackage{multirow}
\usepackage{color}
\usepackage{txfonts}
\usepackage[colorlinks, citecolor=blue, anchorcolor=red,
menucolor=red, linkcolor=red, filecolor=red, urlcolor=blue, frenchlinks=red]{hyperref}
\begin{document}
\title{Is $f_X(1500)$ observed in the $B\to \pi(K)KK$ decays $\rho^0(1450)$? }
\author{Zhi-Tian Zou$^1$} \email{zouzt@ytu.edu.cn}
\author{Ying Li$^{1,2}$} \email{liying@ytu.edu.cn}
\author{Hsiang-nan Li$^3$} \email{hnli@phys.sinica.edu.tw}
\affiliation
{$1$ Department of Physics, Yantai University, Yantai 264005, China\\
 $2$ Center for High Energy Physics, Peking University, Beijing 100871, China\\
 $3$ Institute of Physics, Academia Sinica Taipei, Taiwan 115, Republic of China }
\date{\today}
\begin{abstract}
We suggest that the uncertain state $f_X(1500)$ observed by Belle and BaBar more than a decade ago, which has been viewed as a single scalar or a combination of several even spin resonances, is the vector $\rho^0(1450)$ reported recently by LHCb. Adopting the perturbative QCD approach, we determine the di-kaon distribution amplitudes with the $\rho^0(1450)$ resonance from the LHCb data for the quasi-two-body decays $B^{\pm}\to \pi^{\pm}\rho^0(1450)\to\pi^{\pm}K^+K^-$. It is then shown that the $B^+ \to K^+K^+K^-$ decay spectrum around the invariant mass $M(K^+K^-)\sim 1.5~\rm GeV$ measured by BaBar can be well described by the resonant contribution from $\rho^0(1450)$. The broad structure in the $B^{+}\to K^{+} K_SK_S$ spectrum around the invariant mass $1.5~\rm GeV$ of a $K_SK_S$ pair, which $\rho^0(1450)$ cannot decay into because of Bose-Einstein statistics, can be accounted for by a nonresonant $S$-wave contribution alone. The branching fractions and/or the direct $CP$ asymmetries of the $B^{\pm}\to \pi^{\pm}\rho^0(1450)\to\pi^{\pm}K^+K^-$, $B^{+}\to K^{+}\rho^0(1450)\to K^+K^+K^-$ and $B^{0}\to K^{0}\rho^0(1450)\to K^0K^+K^-$ modes are predicted, which can be tested at the ongoing LHCb and Belle-II experiments. We encourage experimental colleagues to scrutinize our postulation by analyzing relevant data with higher precision.
\end{abstract}
\maketitle
\section{Introduction}
Multi-body $B$ meson decays provide a platform not only for exploring complicated QCD dynamics, but for studying inner structures and properties of exotic states. Along this line, many particles have been identified in Dalitz-plot analyses for these decays, among which a resonance named as $f_X(1500)$ was discovered more than a decade ago, but its structure is not yet clear till now. In this work we will investigate the invariant mass spectra for the three-body $B$ meson decays, in which $f_X(1500)$ was observed, and propose a possible interpretation for it.

As early as 2002, the Belle collaboration performed a simplified analysis of the $B^+ \to K^+K^+K^-$ decay \cite{Abe:2002av}, and found, besides the narrow peak at 1.02 GeV corresponding to the resonance $\phi(1020)$, another broad structure around $1.5~\rm GeV$ in the invariant mass $M(K^+K^-)$ spectrum. This structure was hardly compatible with any known single scalar state, like $f_0(1370)$ or $f_0(1500)$, and the possibility being attributed to a nonresonant contribution or a combination of several resonances could not be excluded. The excess at the $K^+K^-$ invariant mass around 1.5 GeV was then referred to as $f_X(1500)$. Subsequently, BaBar \cite{Aubert:2006nu, Aubert:2007sd} and Belle \cite{Garmash:2004wa, Nakahama:2010nj} identified the similar structure in the $M(K^+K^-)$ spectra of the $B^0\to K_SK^+K^-$ and $B^{\pm}\to K^{\pm}K^+K^-$ decays. BaBar also observed an enhancement around $1.5~\rm GeV$ in the Dalitz plot for the $B^+\to \pi^+K^+K^-$ decay \cite{Aubert:2007xb}. The hypothesis with $f_X(1500)$ being of a scalar type, such as a combination of several resonances $f_0(1370)$, $f_0(1500)$, and $f_0(1710)$, was assumed in the above studies. In particular, Belle concluded that $f_X(1500)$ was best described by a scalar with its mass and width consistent with those of $f_0(1500)$ \cite{Garmash:2004wa}.

It should be noticed that BaBar analyzed the $B^\pm\to \pi^\pm K_SK_S $ decays \cite{Aubert:2008aw} to examine the nature of $f_X(1500)$, but found no evidence of $f_X(1500)$ in the invariant mass $M(K_SK_S)$ spectrum. It suggested that $f_X(1500)$ is either a vector meson or something exotic, because $f_X(1500)$, with an even spin, can decay into the $K_SK_S$ state in principle according to the Bose-Einstein statistics. Afterwards, a peak at $M(K_SK_S)$ between 1.5 and 1.6 GeV in a measure of the $B^0\to K_SK_SK_S$ decay was seen, and described by the interference between the $f_0(1710)$ resonance and a nonresonant component \cite{Lees:2011nf}. BaBar, as reanalyzing a larger data sample for the $B^0 \to   K^0_SK^+K^-$, $B^+ \to K^+K^+K^- $, and $B^+ \to K^+K_SK_S $ decays, continued to regard $f_X(1500)$ as a single $f_0(1500)$ or a combination of $f_0(1500)$, $f_2^{\prime}(1525)$ and $f_0(1710)$ \cite{Lees:2012kxa}. Though the result preferred the latter scenario, it was admitted that the properties of $f_X(1500)$, especially its spin, need to be clarified further. However, a concern on the scalar hypothesis for $f_X(1500)$ remains: the meson $f_0(1500)$ couples more strongly to $\pi\pi$ than to $K\overline K$ \cite{Tanabashi:2018oca}, so an interpretation involving $f_0(1500)$ \cite{Minkowski:2004xf} must explain why there is no strong signal of the $B^\pm\to K^\pm f_0(1500)$ channel in the $B^\pm\to K^\pm \pi^+ \pi^-$ decays \cite{Garmash:2004wa, Aubert:2005ce}. Although $f_X(1500)$ has been viewed as a scalar in many literatures, its nature is controversial so far.

Recently, LHCb reported an enhancement around  $1.5~\rm GeV$ in the invariant mass $M(K^+K^-)$ spectra of the $B^{\pm}\to \pi^{\pm}K^+K^-$ decays, which could be well described by a vector resonance $\rho^0(1450)$ \cite{Aaij:2019qps}. Since the mass of $\rho^0(1450)$ is very close to that of $f_X(1500)$ and it has a rather large width \cite{Tanabashi:2018oca}, we wonder whether $f_X(1500)$ observed by BaBar and Belle in the $B\to K(\pi)K^+ K^- $ decays is $\rho^0(1450)$. If it is, the interpretation for the broad structure in the $M(K^+ K^-)$ spectra does not require a combination of several $f_0$ mesons with narrower widths. Besides, the $B^\pm\to \pi^\pm \rho^0(1450)$ channel has been identified in the $B^\pm\to \pi^\pm \pi^+ \pi^-$ decays with a small finite fit fraction \cite{Aaij:2019jaq,Aaij:2019hzr}. Note that the relative $\rho^0(1450)\to\pi\pi$ and $\rho^0(1450)\to K\overline K$ branching fractions are still uncertain, so our hypothesis is not inconsistent with the above data. According to the Bose-Einstein statistics, a vector meson $\rho^0(1450)$ does not decay into the $K_SK_S$ state. A challenge to our postulation is then how to understand the broad structure in the $M(K_SK_S)$ spectra found by BaBar \cite{Lees:2011nf,Lees:2012kxa}. We will show that it is attributed to a nonresonant $S$-wave contribution, though a small component of scalar resonances cannot be excluded. It is emphasized that the nature of $f_X(1500)$ cannot be determined unambiguously within the current data uncertainty, and we propose a possible scenario here.

To assess the above possibility, we will investigate the resonant contribution to the quasi-two-body  $B\to K(\pi)\rho^0(1450)  \to K(\pi)K^+ K^- $ decays and the nonresonant $S$-wave contribution to the $B\to K_SK_S K$ decay. There are several theoretical approaches to nonleptonic three-body $B$ meson decays available in the literature, such as the factorization approach combined with the heavy meson chiral perturbation theory \cite{Cheng:2007si}, the QCD factorization \cite{Krankl:2015fha}, the perturbative QCD approach (PQCD) \cite{Chen:2002th,Wang:2014ira}, and others based on SU(3) symmetry \cite{Gronau:2005ax}. We will employ the PQCD approach based on the $k_T$ factorization, in which hard emission amplitudes and annihilation ones are calculable without endpoint singularities. It will be demonstrated that our results accommodate well the existed BaBar, Belle and LHCb data for both the $B\to K(\pi)\rho^0(1450)  \to K(\pi)K^+ K^- $ and $B\to KK_SK_S$ decays.

The rest of the paper is organized as follows. In Sec.~\ref{sec:framework}, we specify the inputs of the $S$-wave and $P$-wave di-kaon distribution amplitudes involved in the PQCD framework for nonleptonic three-body $B$ meson decays.  In Sec.~\ref{sec:result}, we present the numerical results and elaborate their physical implications. The branching fractions and the direct $CP$ asymmetries of the quasi-two-body decays $B^{\pm}\to \pi^{\pm}\rho^0(1450)\to\pi^{\pm}K^+K^-$, $B^{+}\to K^{+}\rho^0(1450)\to K^+K^+K^-$ and $B^{0}\to K^{0}\rho^0(1450)\to K^0K^+K^-$ are also predicted for comparison with future data. At last, we summarize this work in Sec.~\ref{summary}.

\section{Framework and Inputs}\label{sec:framework}
In this section, we introduce the PQCD approach to quasi-two-body $B$ meson decays, taking $B^- \to \pi^-\rho^0(1450) \to \pi^-K^+K^- $ as an illustration. The effective weak Hamiltonian for the $b\to d q\bar q$ transition governing the above mode is given by \cite{Buchalla:1995vs}
\begin{eqnarray}\label{effhamil}
\mathcal{H}_{eff}=\frac{G_F}{\sqrt{2}}\left\{V_{ub}V^*_{ud}(C_1O_1+C_2O_2)-V_{tb}V^*_{td}\sum_{i=3}^{10}C_i O_i\right\},
\end{eqnarray}
$V_{IJ}$ being the Cabibbo-Kobayashi-Maskawa (CKM) matrix elements and $G_F$ being the Fermi constant. The explicit expressions for the local four-quark operators $O_i$ ($i = 1, ..., 10$) and their corresponding Wilson coefficients $C_i$ can be found in Ref.~\cite{Buchalla:1995vs}.

In the $B$ meson rest frame for the quasi-two-body $B^- \to \pi^-(K^+K^-) $ decay, the bachelor pion recoils against the collimated kaons. The kaon pair originates from two energetic collinear quarks, and gluons exchanged between them are soft in such a configuration. The interaction between the kaon pair and the pion is power suppressed, so it is reasonable to assume the validity of factorization theorem for this decay. We then write the amplitude as
\begin{eqnarray} \label{amp}
\mathcal{A}\sim \Phi_B \otimes {\cal H}\otimes \Phi_{KK}\otimes \Phi_{\pi},
\end{eqnarray}
where $\otimes$ denotes a convolution in parton momenta. $\Phi_B$ and $\Phi_{\pi}$ are the universal nonpurterbative wave functions of the $B$ and $\pi$ mesons, respectively. The wave function $\Phi_{KK}$, describing how two energetic quarks constitute the kaon pair with certain spin, contain both resonant and nonresonant contributions. In the PQCD framework, a hard kernel $\cal H$ involves sufficiently virtual gluons exchanged between the spectator quark and a quark in the four-quark operator. With intrinsic transverse momenta being kept, a decay amplitude, free of an endpoint singularity, can be calculated perturbatively.  For more details of the PQCD approach, we refer interested readers to Ref.~\cite{Li:2003yj}.

The $B$ meson momentum $p_{B}$, the total momentum of the kaon pair $p$, and the momentum of the bachelor pion $p_3$ are given, in the light-cone coordinates, by
\begin{eqnarray}
&&p_{B}=\frac{m_{B}}{\sqrt 2}(1,1,0_{\rm T}), \nonumber\\
&&p =\frac{m_{B}}{\sqrt2}(1,\eta,0_{\rm T}),  \nonumber\\
&&p_3=\frac{m_{B}}{\sqrt 2}(0,1-\eta,0_{\rm T}),
\end{eqnarray}
with the $B$ meson mass $m_{B}$ and $\eta=\omega^2/m^2_{B}$, $\omega^2=p^2$ being the invariant mass squared of the kaon pair. The momenta of the valence quarks in the $B$ meson, the pion, and the kaon pair are chosen as
\begin{eqnarray}
k_{B}=\left(0,x_B p_B^- ,k_{B \rm T}\right),
k= \left( z p^+,0,k_{\rm T}\right),
k_3=\left(0,x_3p_3^-,k_{3{\rm T}}\right),\label{mom-B-k}
\end{eqnarray}
respectively, where the momentum fractions $x_{B}$, $z$ and $x_3$ run from zero to unity. We define the momenta of the $K$ and $\overline K$ mesons in the kaon pair as
\begin{eqnarray}\label{eq:p1p2}
 p_1&=&\left (\zeta p^+, (1-\zeta)\eta p^+, \sqrt{\zeta(1-\zeta)}\omega,p_{1\rm T} \right ),\nonumber\\
 p_2&=& \left ( (1-\zeta) p^+, \zeta\eta p^+, -\sqrt{\zeta(1-\zeta)}\omega,p_{2\rm T} \right ),
\end{eqnarray}
respectively, with the longitudinal momentum fraction $\zeta$, which obey $p_1+p_2=p$.

The nonperturbative wave functions  $\Phi_B $, $\Phi_{\pi}$ and $\Phi_{KK}$ are the key ingredients for a PQCD study of the quasi-two-body $B^- \to \pi^-\rho^0(1450) \to \pi^-K^+K^- $ decay.  The wave functions $\Phi_B $ and $\Phi_{\pi}$ have been determined to some extent by combining theoretical developments and precise experimental data for two-body nonleptonic decays \cite{Ali:2007ff, Wang:2019msf}. The $P$-wave di-kaon wave function is defined as \cite{Wang:2016rlo,Li:2016tpn,Li:2017mao}
\begin{multline}
\Phi_{KK,P}=\frac{1}{\sqrt{2N_c}}\Big[p\mkern-10.5mu/\phi_v(z,\zeta,\omega^2)+\omega\phi_s(z,\zeta,\omega^2) \\
 +\frac{p\mkern-10.5mu/_1 p\mkern-10.5mu/_2-p\mkern-10.5mu/_2p\mkern-10.5mu/_1}{\omega(2\zeta-1)}\phi_t(z,\zeta,\omega^2)\Big],
\label{Pwave}
\end{multline}
$N_c$ being the number of colors, where
\begin{eqnarray}
&&\phi_v(z,\zeta,\omega^2)=\frac{3{\cal F}_v(\omega^2)}{\sqrt{2N_c}}z(1-z)\left[1+a_vC_2^{3/2}(2z-1)\right]P(\zeta),\nonumber
\end{eqnarray}
and
\begin{eqnarray}
&&\phi_s(z,\zeta,\omega^2)=\frac{3{\cal F}_s(\omega^2)}{2\sqrt{2N_c}}(1-2z)\left[1+a_s(1-10z+10z^2)\right]P(\zeta),\nonumber\\
&&\phi_t(z,\zeta,\omega^2)=\frac{3{\cal F}_t(\omega^2)}{2\sqrt{2N_c}}(2z-1)^2\left[1+a_tC_2^{3/2}(2z-1)\right]P(\zeta),
\end{eqnarray}
are the twist-2 and 3 light-cone distribution amplitudes with the Legendre polynomial $P(\zeta)=2\zeta-1$, respectively. The Gegenbauer moments $a_v$, $a_s$ and $a_t$ associated with the $\rho^0(1450)$ resonance will be fixed later. The form factors ${\cal F}_i(\omega^2)$, $i=v, s, t$, that collect interactions between the two kaons, include both resonant and nonresonant contributions,
\begin{eqnarray}\label{formfactor}
{\cal F}_i(\omega^2)={\cal F}_i^{R}(\omega^2)+{\cal F}_i^{NR}(\omega^2).
\end{eqnarray}
We follow Ref.~\cite{Aaij:2019qps} for the parametrization of ${\cal F}_v^{R}(\omega^2)$, adopting the relativistic Breit-Wigner (RBW) model \cite{Back:2017zqt} with Blatt-Weisskopf barrier factor \cite{blatt},
\begin{eqnarray}
{\cal F}_v^{R}(\omega^2)=\frac{m_{\rho^0(1450)}^2e^{i\beta}}{m_{\rho^0(1450)}^2-\omega^2-i m_{\rho^0(1450)} \Gamma(\omega)}.
\label{fv}
\end{eqnarray}
The nonperturbative strong phase $\beta$ will be set to zero for convenience, though it may affect direct $CP$ asymmetries in some cases. The mass dependent width $\Gamma(\omega)$ is expressed as
\begin{eqnarray}
\Gamma(\omega)=\Gamma_0\left(\frac{q}{q_0}\right)^3\frac{m_{\rho^0(1450)}}{\omega}\frac{X_1(q)}{X_1(q_0)},
\end{eqnarray}
where $\Gamma_0$ is the total width of $\rho^0(1450)$, $q$ is the momentum of either daughter in the rest frame of the resonance, and $q_0$ represents the value of $q$ at $\omega = m_{\rho^0(1450)}$. The Blatt-Weisskopf barrier factor for $J=1$ is given by \cite{Lees:2012kxa}
\begin{eqnarray}
X_1(q)=\sqrt{\frac{1}{1+r^2q^2}},
\end{eqnarray}
with the meson radius parameter $r=4.0$ ${\rm GeV}^{-1}$. We have confirmed that our results are insensitive to the variation of the parameter $r$. As for the form factors $F_{s, t}^{R}(\omega^2)$, we assume the relation \cite{Wang:2016rlo}
\begin{eqnarray}
\frac{{\cal F}_{s, t}^{R}(\omega^2)}{{\cal F}_v^{R}(\omega^2)}=\frac{f_{\rho^0(1450)}^T}{f_{\rho^0(1450)}},
\end{eqnarray}
which will be approximated by $f^T_{\rho(770)} / f_{\rho(770)}$ with $f_{\rho(770)}^T=0.184$ GeV and $f_{\rho(770)}=0.216$ GeV \cite{Wang:2016rlo}. To evaluate the nonresonant contribution, we employ the parameterizations in the whole range of $\omega^2$ \cite{Chen:2002th}
\begin{eqnarray}
{\cal F}_v^{NR}(\omega^2)=\frac{m_P^2}{\omega^2+m_P^2},\,\,\,{\cal F}_{s, t}^{NR}(\omega^2)=\frac{m_{0}m_P^2}{\omega^3+m_0m_P^2},
\end{eqnarray}
where $m_{0}=1.7~{\rm GeV}$ is the chiral mass for a kaon, and the parameter $m_P=1~{\rm GeV}$ has been fixed from fits performed in \cite{Chen:2002th, Chen:2004az}.  It is noticed that the resonant contribution is much larger than the nonresonant one around $\omega\sim 1.45 ~{\rm GeV}$ for the $K^+K^-$ pair, so the term ${\cal F}_i^{NR}(\omega^2)$ in Eq.~(\ref{formfactor}) can be dropped safely. Since we consider only a single resonance $\rho(1450)$ in Eq.~(\ref{fv}), setting the phase $\beta$ to zero is justified.

The $P$-wave contribution to the $K_SK_S$ final state is forbidden by the Bose-Einstein symmetry, so we consider the $S$-wave contribution. The $S$-wave di-kaon wave function has been discussed in \cite{Zou:2020atb}, which takes the form
\begin{multline}
\Phi_{KK,S}=\frac{1}{\sqrt{2Nc}}[P\mkern-11.5mu/\phi_S(z,\zeta,\omega^2)\\+\omega\phi_S^s(z,\zeta,\omega^2)+\omega(n\mkern-9.5mu/ v\mkern-7.5mu/-1)\phi_S^t(z,\zeta,\omega^2)],
\end{multline}
with the light-cone distribution amplitudes
\begin{eqnarray}
\phi_S(z,\zeta,\omega^2)&=&\frac{9}{\sqrt{2Nc}}F_S(\omega^2)a_Sz(1-z)(2z-1),\nonumber\\
\phi_S^s(z,\zeta,\omega^2)&=&\frac{1}{2\sqrt{2Nc}}F_S(\omega^2),\nonumber\\
\phi_S^t(z,\zeta,\omega^2)&=&\frac{1}{2\sqrt{2Nc}}F_S(\omega^2)(1-2z),
\end{eqnarray}
$a_S=-0.5\pm0.1$ being the Gegenbauer moment. The scalar form factor $F_S(\omega^2)$ can also be decomposed into a resonant piece and a nonresonant piece as the Eq.~(\ref{formfactor}), but only the latter contributes here. Hence, we parametrize $F_S(\omega^2)$ as
\begin{eqnarray}
F_S(\omega^2)=\frac{m_S^2}{\omega^2+m_S^2},\label{fs}
\end{eqnarray}
where the parameter is chosen as $m_S=(1.5\pm0.2)~{\rm GeV}$ in the numerical analysis below.

\section{Numerical Results and Discussions}\label{sec:result}
Inputting the above wave functions into the PQCD factorization formula inferred from \cite{Zou:2020atb}, we calculate the amplitude $\cal A$ in Eq.~(\ref{amp}) for the quasi-two-body decay $B^- \to \pi^-\rho^0(1450) \to \pi^-K^+K^- $. The differential decay rate is then given by
\begin{eqnarray}
\frac{d{\cal B}}{d\omega^2}=\tau_B \frac{|\vec p_K| |\vec p_\pi|}{64\pi^3m_B^3} |{\cal A}|^2,
\end{eqnarray}
where $\tau_B$ denotes the $B$ meson lifetime, $|\vec p_K|$ and $|\vec p_\pi|$ are the magnitudes of the kaon and pion momenta in the center-of-mass frame of the kaon pair.
The parameters involved in our numerical study, such as masses, life times and decay widths, are adopted from PDG \cite{Tanabashi:2018oca}.

The Gegenbauer moments $a_{v,s,t}$ in the distributions amplitudes of $\phi_{v,s,t}$ can be derived in nonperturbative methods in principle, such as QCD sum rules and lattice QCD, which are, however, not yet available. LHCb has reported their first amplitude analysis for the $B^{\pm}\to\pi^{\pm}K^+K^-$ decays based on a data sample associated with an integrated luminosity of 3.0 $fb^{-1}$, and found that the data can be well described by a coherent sum of five resonant structures plus a nonresonant component and a contribution caused by $\pi\pi$-$KK$ rescattering \cite{Aaij:2019qps}. As stated in the Introduction, they reported an unexpected $\rho^0(1450)$ resonance, which contributes to the $K^+K^-$ channel with the fit fraction as large as $(30.7\pm1.2\pm0.9)\%$. A detailed theoretical assessment on this result is referred to Ref.~\cite{Wang:2020plx}. Using the $B^{\pm} \to \pi^{\pm}K^+K^-$ branching fraction in PDG \cite{Tanabashi:2018oca}, we deduce the observed branching fraction of the quasi-two-body decay $B^{\pm}\to \pi^{\pm}\rho^0(1450)\to \pi^{\pm}K^+K^-$ as
\begin{align}\label{experesultsLHC}
{\cal B}(B^{\pm}\to \pi^{\pm}\rho^0(1450)\to \pi^{\pm}K^+K^- )= (1.60^{+0.21}_{-0.20} )\times 10^{-6}.
\end{align}
The choice of the Gegenbauer moments
\begin{align} \label{GM}
a_v=-0.70\pm 0.14,\;a_s=-0.50\pm0.10,\,a_t=-0.60\pm0.12,
\end{align}
together with the form factors in Eq.~(\ref{formfactor}), yield the branching fraction consistent with Eq.~(\ref{experesultsLHC}),
\begin{eqnarray}
{\cal B}(B^{\pm}\to \pi^{\pm}\rho^0(1450)\to \pi^{\pm}K^+K^- )= (1.61^{+1.21}_{-0.87} )\times 10^{-6}.\label{pre}
\end{eqnarray}
There are three types of uncertainties in the PQCD framework. The first type is from the initial and final state wave functions, such as the $B$ meson decay constant $f_B=0.19\pm0.02$ GeV and shape parameter $\omega_B=0.4\pm0.04$ GeV in the $B$ meson distribution amplitude,  the Gegenbauer moments in the pion and kaon distribution amplitudes, and the Gegenbauer moments $a_{v,s,t}$ in the di-kaon distribution amplitudes, whose values are varied in a $20\%$ range. The second type comes from the unknown QCD radiative and higher power corrections characterized by the variations of the QCD scale $\Lambda_{QCD}=0.25\pm0.05$ GeV and of the factorization scale $t$ in a $20\%$ range \cite{Li:2012nk,Shen:2018abs}. The last error is caused by the CKM matrix elements. All the above errors have been added in quadrature in Eq.~(\ref{pre}).

Note that the result in Eq.~(\ref{pre}) is larger than the prediction $8.96\times 10^{-8}$ in \cite{Wang:2020plx} by an order of magnitude. We point out the differences between our approach and the one in \cite{Wang:2020plx} and elaborate them below: we focus only on the contribution from $\rho^0(1450)$, but the latter also included a sizable contribution from a virtual $\rho^0(770)$ with a weight factor. The di-pion wave function was inputted directly for $\Phi_{KK}$ based on SU(3) symmetry in \cite{Wang:2020plx}. It turns out that $a_v$ and $a_s$ from our fit have the same sign and order of magnitudes, but $a_t$ has an opposite sign compared to the corresponding Gegenbauer moments in \cite{Wang:2020plx}.

First, the weights associated with the series of $\rho$ resonances, assumed to be real, were fitted to the data for the kaon electromagnetic form factor in \cite{Bruch:2004py}. Hence, it is not legitimate to adopt these real weights, and then vary the relative phases among them arbitrarily to predict the resonant contributions to the $B\to\pi\rho\to\pi KK$ decays. That is, the peak structures corresponding to the nonvanishing relative phases obtained in \cite{Wang:2020plx}, which are nearby the LHCb reported enhancement around the invariant mass 1.5 GeV, are not justified: the same weights with arbitrary relative phases cannot accommodate the kaon form factor data.
The weights associated with a series of resonances are in fact complex numbers to respect the unitarity \cite{Hammoud:2020aqi}. Once the kaon form factor data are refit with the complex weights, the magnitudes of these weights may change, and it is not sure that the $\rho(770)$ contribution will be still comparable to the $\rho(1450)$ one around the invariant mass 1.5 GeV.
Moreover, Eq.~(\ref{experesultsLHC}) is deduced from the LHCb result in \cite{Aaij:2019qps}, where a single $\rho(1450)$ resonance was assumed. To make sense out of the comparison with the LHCb analysis, we have to follow the same assumption, and employ the single resonance parametrization in Eq.~(\ref{fv}), instead of the real-weight multiple-resonance one in \cite{Bruch:2004py}. We stress that our treatment should be taken cautiously: it does not mean that possible contributions from other $\rho$ resonances like $\rho(700)$, $\rho(1250)$,..., have been excluded completely.


We notice that the prefactor $1/2$ in Eq.~(18) of \cite{Wang:2020plx} comes from the definition of the electromagnetic current, which depends on the charges of involved quarks. The currents used to define the di-kaon wave functions arise from the Fierz transformation, which do not contain quark charges. One has to relate the form factors appearing in the di-kaon wave functions to the kaon electromagnetic form factor. For example, the $\bar u\gamma_\mu u$ current for the di-kaon wave functions results in the electromagnetic form factor with the coefficient unity. By correcting the wrong coefficient in \cite{Wang:2020plx}, the predicted $B\to\pi KK$ branching fractions will be enhanced by a factor of 4. Our single-resonance parametrization in Eq.~(\ref{fv}) has the coefficient unity, which is definitely larger than the weight of $\rho(1450)$ in the multiple-resonance parametrization in \cite{Wang:2020plx}. This difference provides further enhancement of the resonant contribution to the $B^{\pm}\to\pi^{\pm}\rho^0(1450)\to\pi^{\pm}K^+K^-$ decays. If we adopt the di-kaon wave functions in \cite{Wang:2020plx}, whose Gegenbauer moments are the same as of the di-pion wave functions actually, the above branching fractions will increase by about 50\%. This is not an order-of-magnitude change, implying that the single-resonance parametrization in Eq.~(\ref{fv}) causes the dominant enhancement in our calculation.

\begin{figure*}[!htbp]
\includegraphics[width=7.50cm]{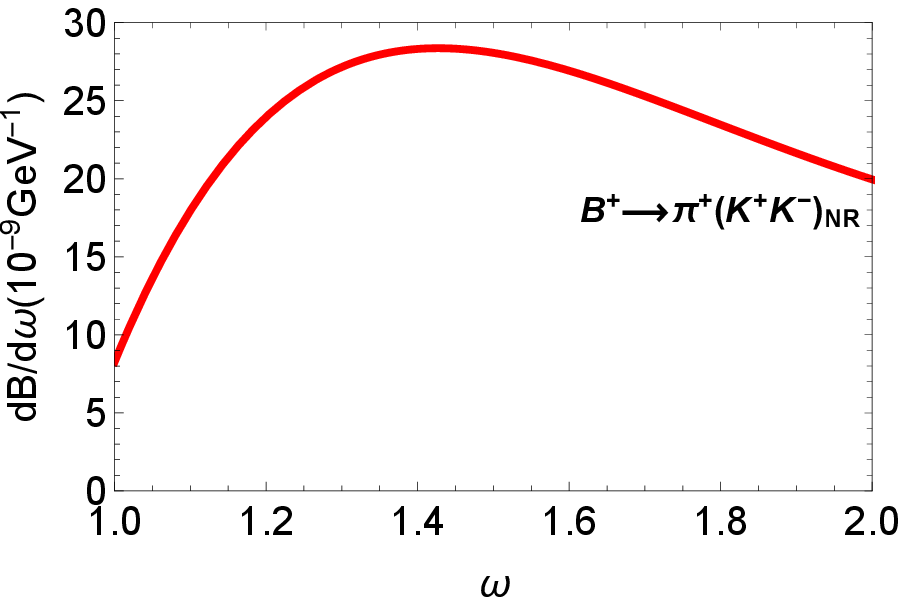}\,\,\,\,
\includegraphics[width=7.50cm]{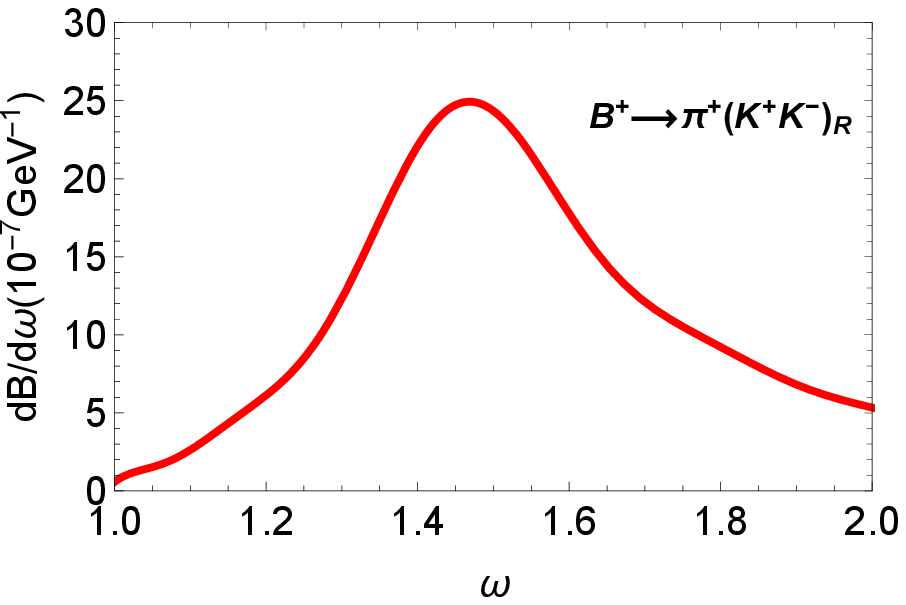}
\put(-340,-20){ (a)}
\put(-110,-20){ (b)}
\caption{$\omega$ dependencies of the $B^+ \to \pi^{+}K^+K^-$ differential branching fractions from (a) the nonresonant contribution and (b) the resonant contribution.}
\label{fig-1}
\end{figure*}

We present in Figs.~\ref{fig-1}(a) and \ref{fig-1}(b) the dependencies of the $B^{+}\to \pi^{+}K^+K^-$ differential branching fractions on the invariant mass $\omega=M(K^+K^-)$ from the nonresonant contribution and from the resonant contribution of $\rho^0(1450)$, respectively. It is seen that the latter is two orders of magnitude larger than the former. The integrated $B^{+}\to \pi^{+}K^+K^-$ branching fraction in Fig.~\ref{fig-1}(a) for the whole kinematic range of $\omega$,
\begin{align}\label{pKKNR}
{\cal B}( B^{\pm}\to \pi^{\pm} K^+K^- )= (6.84^{+0.92}_{-0.87} )\times 10^{-8},
\end{align}
confirms that the nonresonant contribution is negligible.

\begin{figure*}[!htbp]
\includegraphics[width=7.50cm]{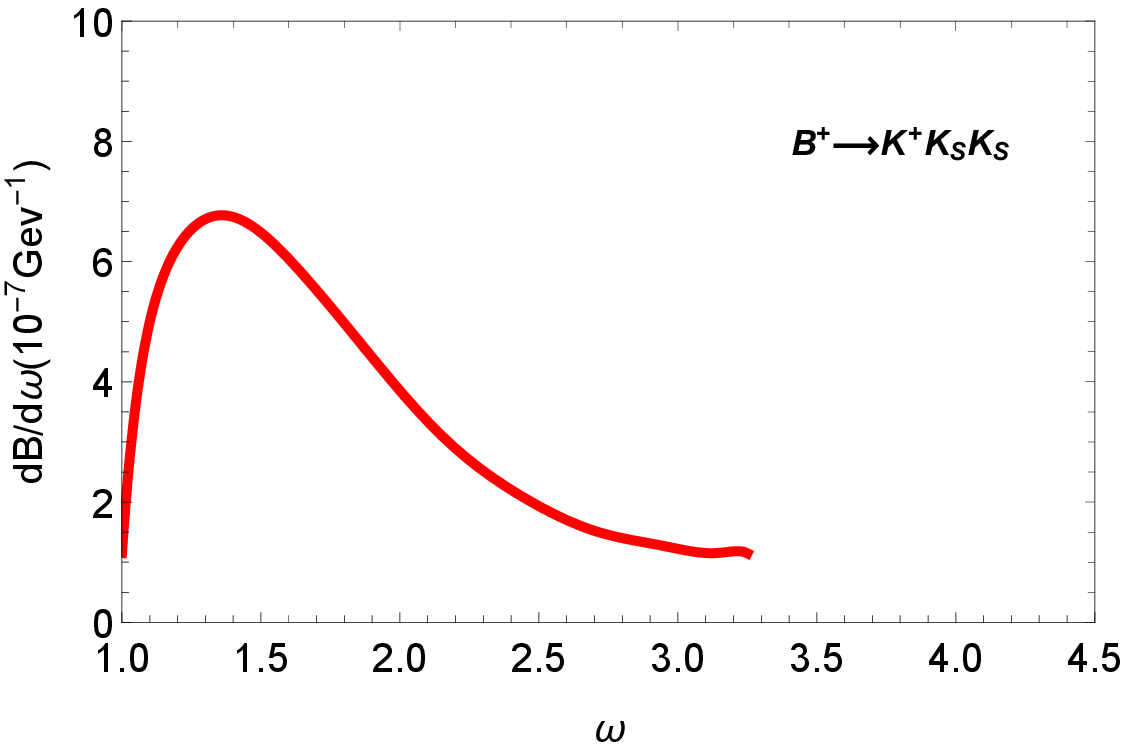}
\includegraphics[width=7.50cm,keepaspectratio]{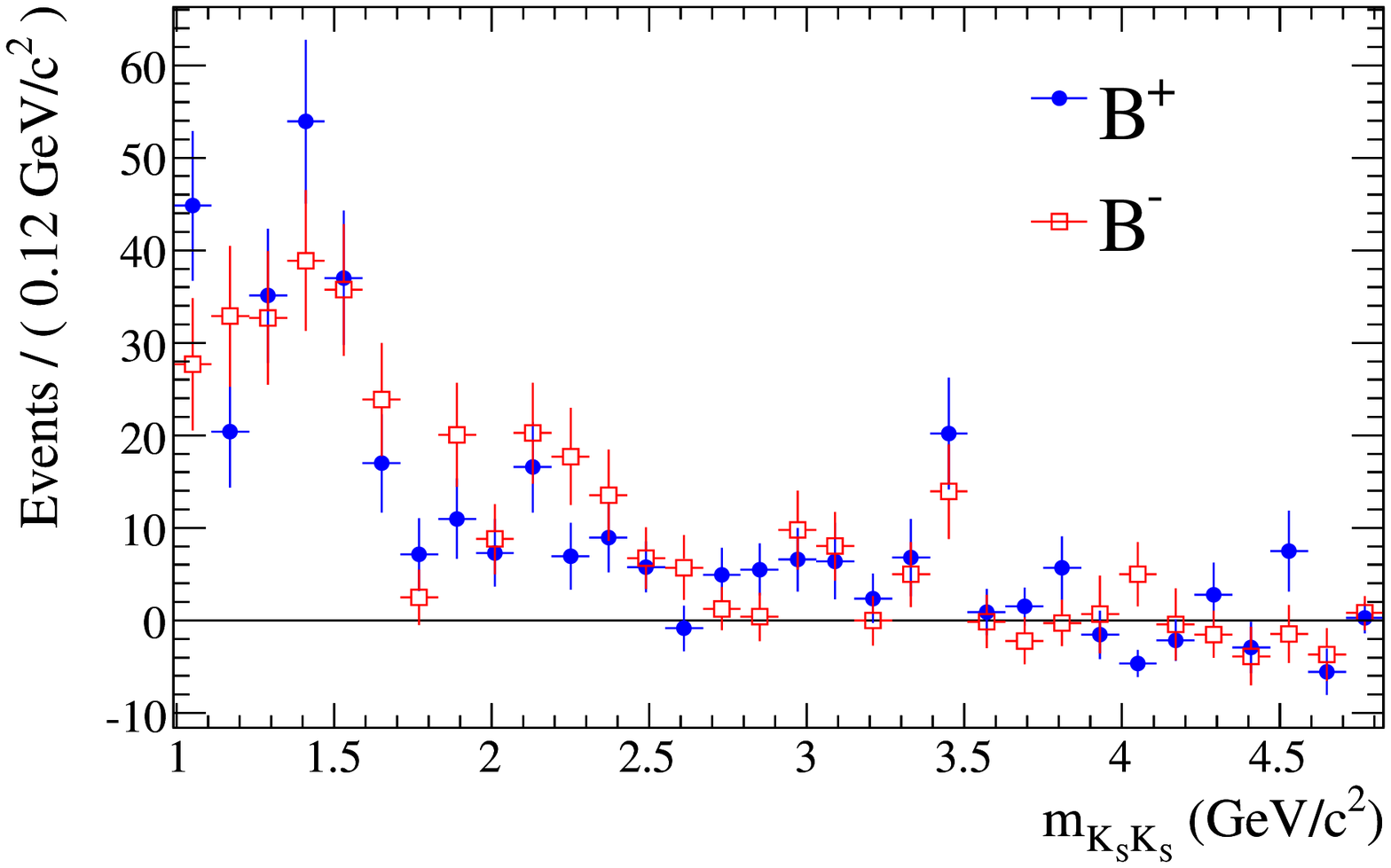}
\put(-340,-20){ (a)}
\put(-110,-20){ (b)}
\caption{(a) predicted $\omega$ dependence of the $B^+ \to  K^{+}K_SK_S$ differential branching fraction, and (b) signal-weighted $m_{K_SK_S}$ distribution of the observed $B^\pm \to  K^{\pm}K_SK_S$ candidates, plotted separately for the $B^+$ and $B^-$ events \cite{Lees:2012kxa}.}
\label{fig-2}
\end{figure*}

As aforementioned, once $f_X(1500)$ is regarded as the vector resonance $\rho^0(1450)$, it cannot be seen in the $B^{\pm} \to \pi^{\pm}K_SK_S$ \cite{Aubert:2008aw} and $B^0\to K_SK_SK_S$ \cite{Lees:2011nf} decays, due to the requirement of the Bose-Einstein statistics.  Hence, we rely on the nonresonant $S$-wave contribution parametrized in Eq.~(\ref{fs}), which is usually sizable, to the $B^+ \to K^+K_S K_S$ decay around the $M(K_SK_S)\sim 1.5~{\rm GeV}$ region. It is found that the predicted curve in Fig.~\ref{fig-2}(a) is in good agreement with the data in Fig.~\ref{fig-2}(b), including the location of the peak, the width, and the magnitude relative to the $K^+K^-$ channel in Fig.~\ref{fig-3} below. Therefore, the confusing peak observed in the $B^+ \to K^+K_SK_S$ decay by BaBar \cite{Lees:2012kxa} is probably not from $f_X(1500)$ but from the nonresonant $S$-wave contribution. Considering the large uncertainty of the $B^+\to K^+K_SK_S$ signals, we cannot exclude the possibility of a small component of scalar resonances in the measured $M(K_SK_S)$ spectrum. We hope that experimentalists can collect data with higher precision, so as to confirm the nature of the peak at $m_{K_SK_S}\sim 1.5~{\rm GeV}$ in Fig.~\ref{fig-2}(b).

\begin{figure*}[!htbp]
\includegraphics[width=7.50cm]{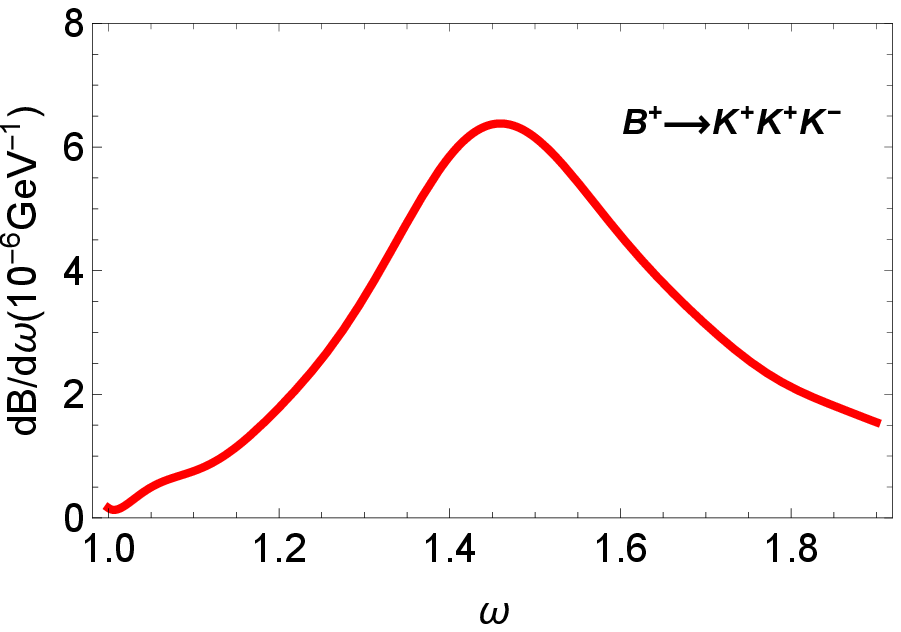}
\includegraphics[width=7.50cm,keepaspectratio]{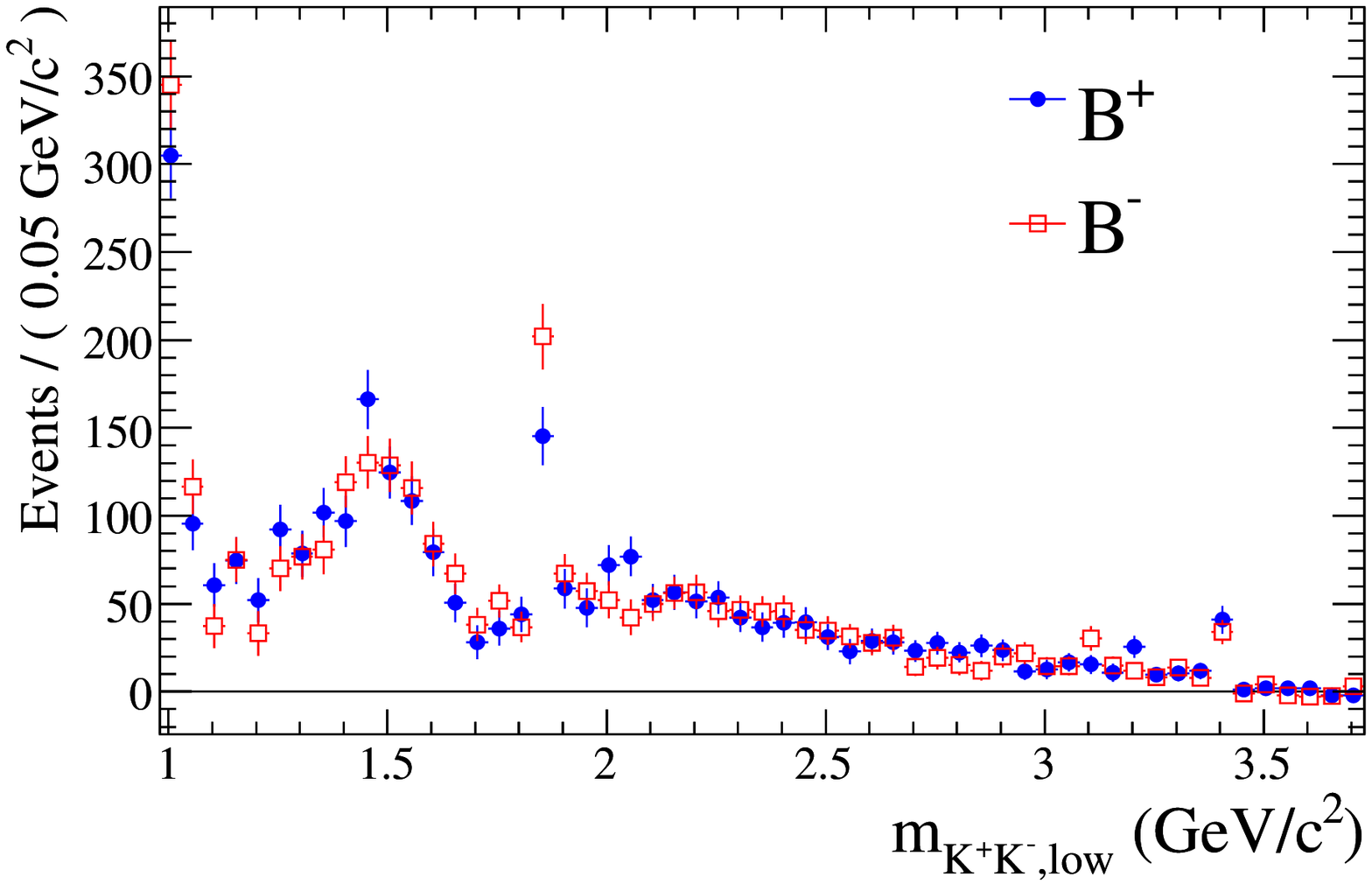}
\put(-340,-20){ (a)}
\put(-110,-20){ (b)}
\caption{(a) predicted $\omega$ dependence of the $B^+ \to  K^{+}K^+K^-$ differential branching fraction, and (b) signal-weighted $m_{K^+K^-}$ distribution for the observed $B^\pm \to  K^{\pm}K^+K^-$ candidates, plotted separately for the $B^+$ and $B^-$ events \cite{Lees:2012kxa}. }\label{fig-3}
\end{figure*}

Next we analyze the quasi-two-body decay $B^+ \to K^{+}\rho^0(1450)\to K^{+}K^+K^-$ with the parameters in Eq.~(\ref{GM}) fitted from the LHCb data, considering only the $\rho^0(1450)$ contribution due to the smallness of the nonresonant one. We have checked that the latter amounts only up to 4\% of the former. The predicted differential branching fraction in the invariant mass $M(K^+K^-)$ is displayed in Fig.\ref{fig-3}(a). For comparison, we show the signal-weighted $m_{K^+K^-}$ distributions of the $B^+$ and $B^-$ events from the $B^\pm \to K^\pm K^+ K^-$ decays \cite{Lees:2012kxa} in Fig.~\ref{fig-3}(b), where the enhancement located at $1~{\rm GeV}$ is attributed to $\phi(1020)$. It is obvious that the location of the peak, the width, and the 10 times larger magnitude relative to the $K_SK_S$ channel of our prediction match well the data. That it, the puzzling $f_X(1500)$ structure can be described by a single vector resonance $\rho^0(1450)$.

Supposing $f_X(1500)$ to be $\rho^0(1450)$, we predict the branching fractions of the quasi-two-body $B^{+/0}\to K^{+/0}\rho^0(1450)\to K^{+/0}K^+K^-$ decays in the PQCD approach,
\begin{multline}
 \mathcal{B}(B^+\to K^+\rho^0(1450)\to K^+K^+K^-) \\
 =(3.62^{+1.33+0.87+0.18}_{-1.42-0.81-0.24})\times10^{-6},\nonumber
\end{multline}
\begin{multline}
 \mathcal{B}(B^0\to K^0\rho^0(1450)\to K^0K^+K^-) \\
 =(7.49^{+4.36+3.76+0.35}_{-3.96-3.05-0.12})\times10^{-7},\label{br}
\end{multline}
where the dominant uncertainties arise from the shape parameter of the $B$ meson distribution amplitude and the Gegenbauer moments in Eq.~(\ref{GM}). These results can be confronted with data from the ongoing LHCb and Belle-II experiments.

We also compute the direct $CP$ asymmetries of the above decays, obtaining
\begin{eqnarray}
&\mathcal{A}_{CP}(B^{\pm}\to \pi^{\pm}\rho^0(1450)\to\pi^{\pm}K^+K^-)=(1.31^{+39.20}_{-10.10})\%,\nonumber\\
&\mathcal{A}_{CP}(B^{+}\to K^{+}\rho^0(1450)\to K^+K^+K^-)=(3.73^{+10.77}_{-5.57})\%,\nonumber\\
&\mathcal{A}_{CP}(B^{0}\to K^{0}\rho^0(1450)\to K^0K^+K^-)=(35.9_{-47.0}^{+25.1})\%.
\end{eqnarray}
The quasi-two-body decays $B^{\pm}\to \pi^{\pm}\rho^0(1450)\to\pi^{\pm}K^+K^-$ are the CKM favored color-allowed tree dominant processes, and the QCD penguin contributions to the $u\bar u$ and $d\bar d$ components of $\rho^0(1450)$ cancel each other. The small direct $CP$ asymmetry in these decays is then understood, which is proportional to the interference between the tree and penguin contributions. LHCb reported $(-10.9\pm4.4\pm2.4)\%$ for this direct $CP$ asymmetry \cite{Aaij:2019qps}, whose uncertainty is still large. BaBar has observed the direct $CP$ asymmetry $(-6\pm28\pm20^{+12}_{-35})\%$ \cite{Aubert:2009av} in the $B^{\pm}\to \pi^{\pm}\rho^0(1450)\to \pi^{\pm}\pi^+\pi^-$ decays, and $(-6\pm28)\%$ in the corresponding two-body decay $B^{\pm}\to\pi^{\pm}\rho^0(1450)$ under the narrow width approximation. Along the same line, we get the direct $CP$ asymmetry about $1.3\%$ for $B^{\pm}\to\pi^{\pm}\rho^0(1450)$ through the quasi-two-body decays $B^{\pm}\to\pi^{\pm}\rho^0(1450)\to\pi^{\pm}K^+K^-$, in agreement with the BaBar measurement. Recently,  LHCb measured the direct $CP$ asymmetry of $B^{\pm}\to \pi^{\pm}\rho^0(1450)\to \pi^{\pm}\pi^+\pi^-$ based on three different models in \cite{Aaij:2019hzr, Aaij:2019jaq}, and the data are consistent with zero but with different signs. On the contrary, the $B^{+}\to K^{+}\rho^0(1450)\to K^+K^+K^-$ decay is a penguin dominant process, to which the tree  contribution is small, so its direct $CP$ asymmetry is only few percent. As for another penguin dominated mode $B^{0}\to K^{0}\rho^0(1450)\to K^0K^+K^-$, the contribution from the QCD penguin is cancelled, such that the tree and penguin contributions are comparable, leading to a larger direct $CP$ asymmetry. It is also the reason why its branching fraction is much smaller than that of the $B^{+}\to K^{+}\rho^0(1450)\to K^+K^+K^-$ decay as indicated in Eq.~(\ref{br}). We remind that the $\pi\pi$-$KK$ rescattering may affect direct $CP$ asymmetries remarkably, and more precise data will uncover its impact.
\section{Summary}\label{summary}
In this paper we have examined whether  the puzzling $f_X(1500)$ that has been modeled as a single scalar or a combination of several scalar resonances by BaBar and Belle for more than a decade is the vector $\rho^0(1450)$ reported by LHCb recently. If it is, $\rho^0(1450)$ with its large width can accommodate the broad enhancement around $1.5~\rm GeV$ in the invariant mass $M(K^+K^-)$ spectra of the $B\to K(\pi)K^+K^- $ decays naturally. To meet the same purpose, one usually needs several scalar resonances with narrower widths. Moreover, no strong signal of $B^\pm\to K^\pm f_0(1500)$ in the $B^\pm\to K^\pm \pi^+ \pi^-$ decays can be explained easily. Our hypothesis is not inconsistent with the small fit fraction of $B^\pm\to \pi^\pm \rho^0(1450)$ in the $B^\pm\to \pi^\pm \pi^+ \pi^-$ decays, because the relative $\rho^0(1450)\to\pi\pi$ and $\rho^0(1450)\to K\overline K$ branching fractions are still uncertain. Since $\rho^0(1450)$ cannot be seen in the  $B^0\to K_S K_SK_S$ decay, we have attributed the broad structure in the $M(K_SK_S)$ spectrum identified by BaBar to a nonresonant $S$-wave contribution, which fits the observed feature of a wide peak with its height lower than from a resonant contribution.

To verify the above hypothesis quantitatively, we have studied the relevant three-body $B$ meson decays in the PQCD approach. We determined the di-kaon distribution amplitudes from the LHCb data for the $B^{\pm}\to \pi^{\pm}K^+K^-$ decays, and then calculated the resonant contribution to the quasi-two-body decays $B^{+}\to K^{+}\rho^0(1450)\to K^+K^+K^-$ and the nonresonant $S$-wave contribution to the $B^{+}\to K^{+} K_SK_S$ decay. The obtained differential branching fractions agree well with the experimental data for both modes in the locations of the peaks, the widths, and the relative magnitudes between them around the invariant mass 1.5 GeV.  We acknowledge that possible contributions from other $\rho$ resonances like $\rho(700)$, $\rho(1250)$,... have not been excluded completely, though the single resonance parametrization was assumed in this work. It should be also stressed that a small component of scalar resonances in the $M(K_SK_S)$ spectrum cannot be excluded within the current data uncertainty. We suggest experimental colleagues to collect more precise data for the $B^{+}\to K^{+} K_SK_S$ decay, so as to scrutinize our postulation and clarify the nature of the peak at $M(K_SK_S)\sim 1.5~{\rm GeV}$. The branching fractions and/or the direct $CP$ asymmetries of the quasi-two-body decays $B^{\pm}\to \pi^{\pm}\rho^0(1450)\to\pi^{\pm}K^+K^-$, $B^{+}\to K^{+}\rho^0(1450)\to K^+K^+K^-$ and $B^{0}\to K^{0}\rho^0(1450)\to K^0K^+K^-$ have been also predicted, which can be tested at the ongoing LHCb and Belle-II experiments.

\section*{Acknowledgment}
We warmly thank H.Y. Cheng for constructive discussions and reading the manuscript carefully. This work was supported in part by the National Science Foundation of China under the Grant Nos.~11705159 and 11975195, by the Natural Science Foundation of Shandong province under the Grant No. ZR2018JL001 and No.ZR2019JQ04, by the Project of Shandong Province Higher Educational Science and Technology Program under Grants No. 2019KJJ007, and by MOST of R.O.C. under Grant No. MOST-107-2119-M-001-035-MY3. Z.T. Zou acknowledges Institute of Physics, Academia Sinica for their hospitality to his visit, during which part of the work was done.

\end{document}